\documentclass[rnote]{aa}
\usepackage{graphicx}
\usepackage{txfonts}
\usepackage{verbatim}
\usepackage{savesym}
\savesymbol{iint}
\savesymbol{iiint}
\savesymbol{iiiint}
\savesymbol{idotsint}
\usepackage{amsmath}
\usepackage{natbib}
\newcommand\fy{\ensuremath{\overset{\text{y}}{.}}}
\newcommand{\FINAL}{} 

\begin{document}
\title{Periodicity in some light curves of the solar analogue V352~CMa
\thanks{The analysed photometry and numerical results of the analysis
are both published electronically at the CDS via anonymous ftp to
cdsarc.u-strasbg.fr (130.79.128.5) or via 
http://cdsarc.u-strasbg.fr/viz-bin/qcat?J/A+A/yyy/Axxx}}
\author{P. Kajatkari      \inst{1}
\and L. Jetsu             \inst{1}
\and E. Cole              \inst{1}
\and T. Hackman           \inst{1,2}
\and G.W. Henry           \inst{3}
\and S-L. Joutsiniemi     \inst{1}
\and J. Lehtinen          \inst{1}
\and V. M\"akel\"a        \inst{1}
\and S. Porceddu          \inst{1}
\and K. Ryyn\"anen        \inst{1}
\and V. \c{S}olea             \inst{1}}
\institute{Department of Physics, Gustaf H\"{a}llstr\"{o}min katu 2a 
(P.O. Box 64), FI-00014 University of Helsinki, Finland
\and
Finnish Centre for Astronomy with ESO (FINCA), University of Turku, 
V\"ais\"al\"antie 20, FI-21500 Piikki\"o, Finland
\and
Center of Excellence in Information Systems, Tennessee State University, 
3500 John A. Merritt Blvd., Box 9501, Nashville, TN 37209, USA}
\date{Received / Accepted}
\abstract{}{Our aim was to study the light curve periodicity of the solar analogue \object{V352~CMa} and in particular show that the presence or absence of periodicity in low amplitude light curves can be modelled with the Continuous Period Search (CPS) method.}
{We applied the CPS method to 14 years of V-band photometry of V352 CMa and obtained estimates for the mean, amplitude, period and minima of the light curves in the selected datasets. We also applied the Power Spectrum Method (PSM) to these datasets and compared the performance of this frequently applied method to that of CPS. }
{We detected signs of a $11.7 \pm 0.5$ year cycle in 
in the mean brightness. 
The long--term average
photometric rotation period was $7.24 \pm 0.22$ days. 
The lower limit for the
differential rotation coefficient would be $|k|>0.12$, 
if the law of solar surface differential
rotation were valid for \object{V352~CMa} and the period changes traced this phenomenon.
Signs of stable active longitudes rotating with 
a period of $7.157 \pm 0.002$ days were
detected from the epochs of the light minima with the Kuiper method. 
CPS performed better than the traditional PSM, because
the latter always used a sinusoidal model for the data 
even when this was clearly not the correct model.}{}

\maketitle

\section{Introduction \label{intro}}

\object{V352~CMa} (\object{HD~43162}) was among
the first 384 bright extreme ultraviolet sources 
detected by the ROSAT satellite \citep{Sha93}.
Metallicity and space motion indicated
that it belongs to the young disk population \citep{Egg95}.
It was included into the list of 38 nearby young 
single solar analogs having ages between 0.2 and 0.8 Gyr \citep{Gai98}.
We collected some of its physical parameters into Table \ref{physparam}. 
Strong Ca H\&K emission ($R'_{\mathrm{HK}}$), 
high X-ray luminosity ($R_{\mathrm{X}}$) 
and 
rapid rotation ($v \sin{i}$)
indicate that it is a young star. 
The high lithium abundance 
also supports this \citep{Gai98,San04a}.
However, 
the age and metallicity ([Fe/H]) estimates 
are inconsistent (Table \ref{physparam}).
Except for \cite{Gai00} and \cite{Gai02},
the effective surface temperature 
($T_{\mathrm{eff}}$) and 
gravity ($\log{g}$)
estimates agree and 
support the spectral type G6.5~V \citep{Gra06}.

\object{V352~CMa} was classified as a member of a 
stellar kinematical group (hereafter SKG) of 19 stars by \cite{Jef93}.
Later, it has been identified as a member another group IC~2391.
This more recently identified SKG contains 29 stars.
The estimated age of IC~2391 is 45~Myr \citep{Nak10,Mal10,Nak12}.
However, only two stars in IC~2391 were among the 19 members of the original SKG defined by \citet{Jef93}. These two stars are 
\object{V352~CMa} and \object{LQ~Hya}. This indicates how difficult it is to confirm the membership in any particular SKG. \cite{Chr03b} observed an extreme ultraviolet flare in a M3.5~V star (\object{EUVE J0613--23.9B}) located 2\farcm5 away from \object{V352~CMa} (\object{EUVE J0613--23.9}). 
We noticed that this M3.5~V star is the same object that was later identified as the binary companion of \object{V352~CMa} \citep{Rag10}.
About 90\% of the ROSAT source positions were within 1\arcmin, and 100\% were within 2\farcm1, of the catalogued positions \citep{Sha93}. 
The 2\farcm5 distance between \object{V352~CMa} and its M3.5 companion is close to the latter 2\farcm1 limit. This companion had no influence to our photometry, because the diameter of the focal-plane diaphragm is only 55\arcsec.

\object{V352~CMa} was in the sample of 11 new stars, 
where a debris disk was detected
using the Spitzer Space Telescope data \citep{Kos09}. 
However, it was the only star in this sample without a known planet. \cite{Cut99} determined the photometric rotation 
period of \object{V352~CMa}, $P_{\mathrm{phot}}=7\fd2 \pm 0\fd2$,
from a light curve with a low $0.03$ peak to peak amplitude.
\cite{Gai00} detected no periodicity in their photometry, 
which were the same data as our first season of photometry. Hipparcos photometry revealed no periodicity
\citep[][n=228 observations]{Koe02}.
\cite{Wri04} derived $P_{\mathrm{phot}}\approx 8^{\mathrm{d}}$
from the observed $B-V$ and $R'_{HK}$ values. 

In this research note, we analyse long-term photometry 
of \object{V352~CMa} with CPS \citep{Leh11}. 
The previous studies have indicated that periodicity may be 
present or absent in any part of this photometry. 
We will show that CPS is an ideal method 
for analysing this type of challenging data.
We also compare CPS to PSM formulated by
\citet{Hor86}.

\addtolength{\tabcolsep}{-0.12cm} 
\begin{table*}
\caption{Physical parameters of \object{V352~CMa}}
\begin{center}
\begin{tabular}{ccccccccrccccccc}
\hline
\hline
$T_{\mathrm{eff}} $&          &$\log{g}   $&                &$v_{\mathrm{rad}}$   &              &
$v \sin{i} $&              &[Fe/H]      &                & Age          &              &
$R'_{\mathrm{HK}} $&              &$R_{\mathrm{X}}    $&                \\
${\mathrm{[K]}} $& Ref          &${\mathrm{[cm~s^{-2}]}}$ & Ref       &${\mathrm{[km~s^{-1}]}}$& Ref      &
${\mathrm{[km~s^{-1}]}}$& Ref    &              & Ref            &[Myr]         & Ref          &
$          $& Ref          &$            $& Ref            \\
\hline
$5480      $& 7            &$ 4.57       $& 10       &$22.7 \pm0.7 $& 1         &
$6\pm2     $& 6            &$-0.15       $& 4        &$200-800$     & 5         &
$-4.40     $& 7            &$-4.26       $& 5        \\
$5593      $& 8            &$ 4.10       $& 11       &$21.69\pm0.16$& 2         &
$ 5.7      $& 7            &$-0.16       $& 9        &$70-800      $& 7         &
$-4.40     $& 12           &$-4.31       $& 7        \\
$5630      $& 10           &$ 4.48       $& 13       &$22.6 \pm0.6 $& 3         &
$ 5.49     $& 13           &$-0.02       $& 10       &$575         $& 12        &
$-4.48     $& 18           &$-4.29       $& 21            \\
$5473      $& 11           &$ 4.49       $& 18       &$21.3-21.8   $& 6         & 
$ 9.63     $& 22           &$-0.11       $& 11       &$<1000       $& 14        &
$-4.39     $& 21           &$-4.39       $& 22  \\
$5633      $& 13           &$ 4.52       $& 23       &$21.7        $& 7         &
$          $&              &$-0.01       $& 13       &$<6500       $& 19        &
$          $&              &$            $&               \\
$5585      $&  15          &$            $&          &$21.9 \pm0.2 $&  16       &
$          $&              &$-0.10       $&  15      &$33-35,280   $&  21       &
$          $&              &$            $&               \\
$5571      $& 18           &$            $&          &$21.91\pm0.09$&  17       &
$          $&              &$-0.10       $&  19      &$5286        $&  24       &
$          $&              &$            $&               \\
$5584      $&  19          &$            $&          &$22.23\pm0.09$&  21       &
$          $&              &$-0.04       $&  20      &$<45         $&  25       &
$          $&              &$            $&               \\
$5590      $&  23          &$            $&         &$            $&            & 
$          $&              &$ 0.02       $&  23     &$            $&            &
$          $&              &$            $&               \\
$          $&              &$            $&         &$            $&            &
$          $&              &$ 0.02       $&  24     &$            $&            &
$          $&              &$            $&               \\
\hline
\end{tabular}
\tablebib{
(1) \cite{Bea81};
(2) \cite{And85};
(3) \cite{Bea86};
(4) \cite{Egg95};
(5) \cite{Gai98};
(6) \cite{Cut99};
(7) \cite{Gai00};
(8) \cite{Dec00};
(9) \cite{Hay01};
(10) \cite{San01};
(11) \cite{Gai02};
(12) \cite{Wri04};
(13) \cite{San04a};
(14) \cite{San04c};
(15) \cite{Nor04};
(16) \cite{Gon06};
(17) \cite{Abt06};
(18) \cite{Gra06};
(19) \cite{Hol09};
(20) \cite{Arn10};
(21) \cite{Mal10};
(22) \cite{Mar10};
(23) \cite{Bru11};
(24) \cite{Fer11};
(25) \cite{Nak12}}
\end{center}
\addtolength{\tabcolsep}{+0.12cm} 
\label{physparam}
\end{table*}

\begin{figure} 
\resizebox{\hsize}{!}{\includegraphics{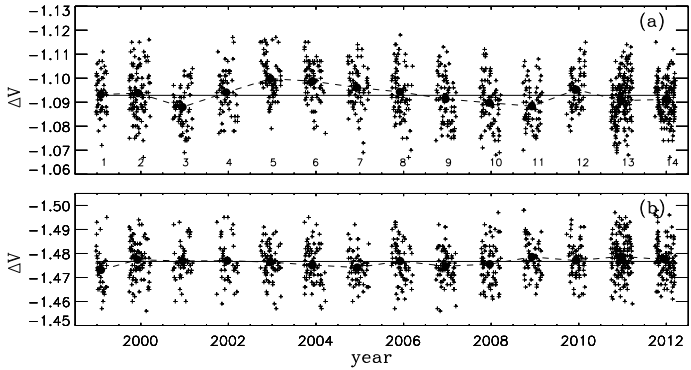}}
\caption{Differential photometry.
{\bf (a)} All $\Delta V_{\mathrm{S-C}}$ data (crosses),
their mean level (continuous line) 
and the seasonal $\Delta V_{\mathrm{S-C}}$ means (filled circles) connected with dashed lines. 
{\bf (b)} All $\Delta V_{\mathrm{K-C}}$ data. The magnitude scale and the notations 
are as in (a)}
\label{figureone}
\end{figure}

\section{Observations}

The photometry of our target star S=\object{V352~CMa} was 
obtained with the automated T3 0.4~m
photoelectric telescope (APT) at Fairborn Observatory in Arizona.
The observations were made between 
December 23rd, 1998 (HJD=2451170.8) \FINAL
and 
March 17th, 2012 (HJD=2456003.6)    \FINAL
during 14 \FINAL 
observing seasons.       
The comparison star was 
C=\object{HD~43879} (F5V, $V=7.26$), \FINAL 
which is a visual double with $V=7.3$ 
and $11.2$ components. 
Both components are included in the 55\arcsec 
focal-plane diaphragm of the photometer, 
because their angular separation is only 6.7\arcsec.
The check star was 
K=\object{HD~43429} (K1~III, $V=5.99$). \FINAL
The standard Johnson differential magnitudes, $\Delta V_{\mathrm{S-C}}$, 
are shown in Fig. \ref{figureone}a.
The mean ($m$) and the standard deviation ($s$) of these $n=1257$
observations were 
$m = -1.0929$ and $s = 0.0096$. \FINAL
The lower panel displays the $\Delta V_{\mathrm{K-C}}$ differential magnitudes.
The standard deviation of these  
$\Delta V_{\mathrm{K-C}}$ observations ($n=1027$) \FINAL 
gave an internal accuracy estimate of 
$0.0079$. \FINAL
We used this accuracy estimate to test the constant brightness hypothesis for  
$\Delta V_{\mathrm{S-C}}$. 
The result, 
$\chi^2=1854$, \FINAL for these $\Delta V_{\mathrm{S-C}}$ observations ($n=1257$)  \FINAL
indicated that the brightness changes of \object{V352~CMa} were real. 
There were also clear trends in the $n=14$ seasonal $\Delta V_{\mathrm{S-C}}$
means having $s=0.0036$ \FINAL
(Fig. \ref{figureone}a: filled circles).
The respective seasonal $\Delta V_{\mathrm{K-C}}$ means were more stable
with $s=0.0016$ \FINAL
(Fig. \ref{figureone}b: filled circles).
More detailed information of the operation of the T3 0.4~m APT 
and the data reduction procedures can be found, 
e.g. in \cite{Hen99} and \cite{Fek05}. 

\begin{figure*} 
\resizebox{\hsize}{!}{\includegraphics{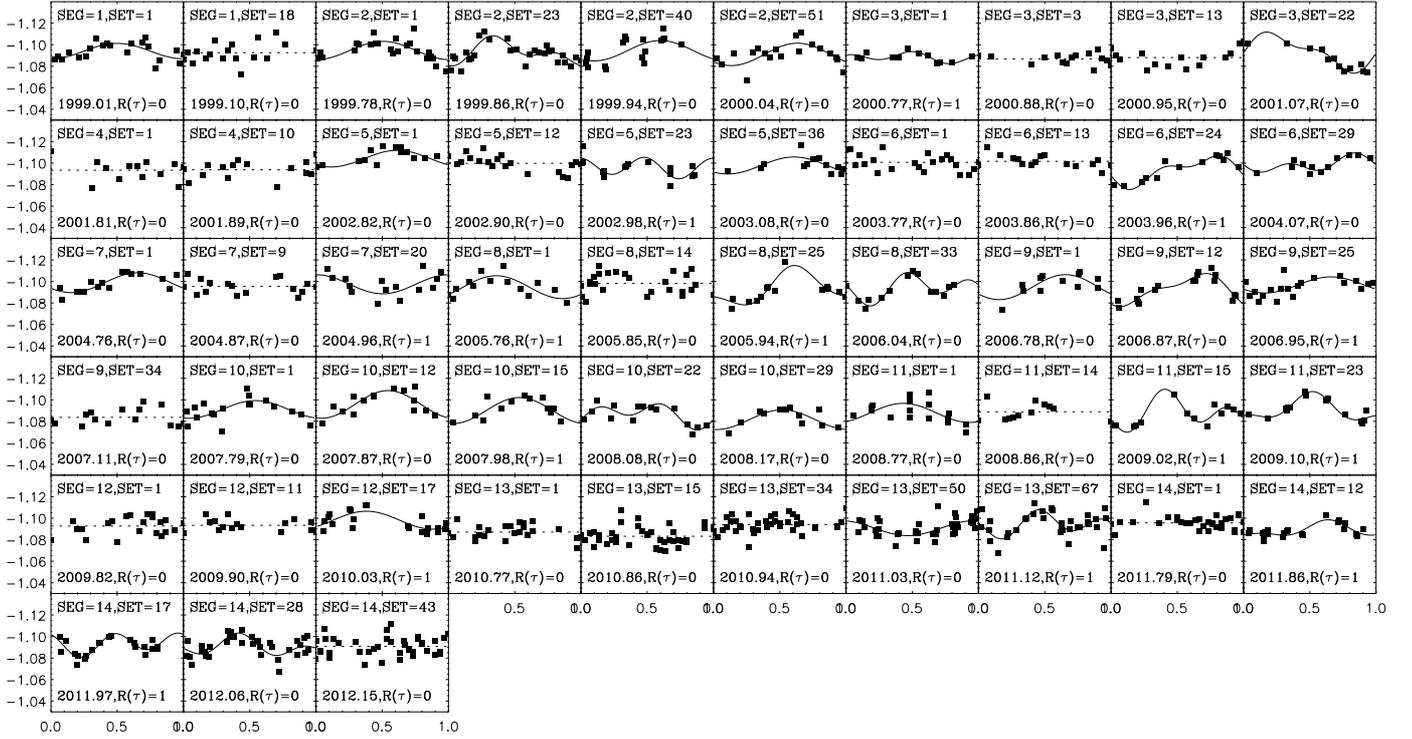}}
\caption{The data and light curves of all independent datasets ($IND(\tau)=1$).
All notations are explained in the 5th paragraph of Sect. \ref{results}}
\label{figuretwo}
\end{figure*}

\section{CPS results \label{results}}

CPS divided the data into segments (SEG) and datasets (SET).
There were 14 segments (i.e. seasons). 
The maximum length of a dataset was $\Delta T_{\rm max}=30^{\mathrm{d}}$.
Each dataset had to contain at least $n \ge n_{\rm min}=12$ 
observations $y_i=\Delta V_{\mathrm{S-C}}(t_i)$. The CPS results correlate for temporally overlapping datasets,
because they contain common data.
To eliminate such correlation, 
we selected the  sequence of independent datasets with no common data. 
The notations ${\rm IND}(\tau)=1$ and 0 are used for independent and not
independent (i.e. overlapping) datasets, respectively.
Our notation for the mean of the $n$ observing times $t_i$ 
of a dataset is $\tau$.

The CPS model is a $K$:th order Fourier series  
\begin{eqnarray}
\hat{y}(t_i) = 
\hat{y}(t_i,\bar{\beta}) = 
M + \sum_{k=1}^K{[B_k\cos{(k2\pi ft_i)} + C_k\sin{(k2\pi ft_i)}]}.
\nonumber
\label{model}
\end{eqnarray}
Of the free parameters $\bar{\beta}=[M,B_1,..,B_K,C_1,...,C_K,f]$,
the physically meaningful ones are
the mean, $M(\tau)$ and the period, $P(\tau)=f^{-1}(\tau)$.
The other free parameters $B_1, ..., B_{\rm K}, C_1, ..., C_K$ give
the total amplitude of the model, $A(\tau)$, 
as well as the epochs of the primary and secondary minima in time,
$t_{\rm min,1}(\tau)$ and $t_{\rm min,2}(\tau)$.
CPS uses the Bayesian information criterion to choose the best modelling order $K$ for each dataset. 
We tested orders $0 \le K \le 2$.  
The total number of modelled datasets was 485.
Of these, no periodicity was detected in 178 datasets ($K=0$), 
where the best model was constant brightness $\hat{y}(t_i)=M(\tau)$.
Periodicity was detected in 307 datasets.
The best orders were $K=1$ in 170 datasets and $K=2$ in 137 datasets.

\begin{figure}
\resizebox{\hsize}{!}{\includegraphics{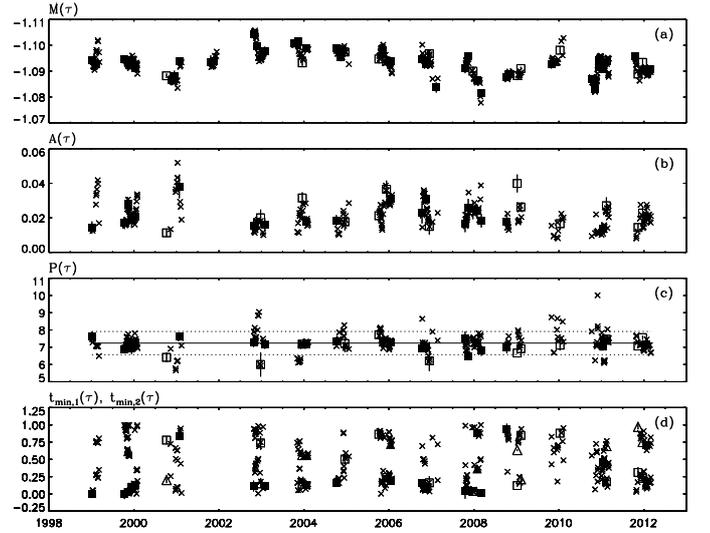}}
\caption{
{\bf (a)} mean $M(\tau)$, 
{\bf (b)} amplitude $A(\tau)$, 
{\bf (c)} period $P(\tau)$
and 
{\bf (d)} phases of the 
primary $t_{\mathrm{min,1}}(\tau)$ and secondary $t_{\mathrm{min,2}}(\tau)$ minima.
The symbols are explained in the 3rd paragraph of Sect. \ref{results}.
The horizontal lines in (c) denote the  $P_{\mathrm{w}} \pm 3 \Delta P_{\mathrm{w}}$ level.
The phases in (d) were calculated from HJD~$2451178.1245+7\fd158$E.}
\label{figurethree}
\end{figure}

The error estimates for $M(\tau)$, $A(\tau)$, $P(\tau)$, 
$t_{\rm min,1}(\tau)$ and $t_{\rm min,2}(\tau)$ were determined 
with the bootstrap method. 
Bootstrap was also used to identify the
reliable or unreliable light curve parameter estimates.
Our notations are $R(\tau)=0=$ reliable estimate 
and $R(\tau)=1=$ unreliable estimate.
The CPS results are published only electronically.
We use the format specified in \cite{Leh11}, 
where a detailed description of CPS can be found.
The CPS results for different parameters are summarised below.
\newcommand{\risti}{{\scriptsize $[\times]$}}
\newcommand{\anelio}{{$[\blacksquare]$}}
\newcommand{\bnelio}{{$[\Box]$}}
\newcommand{\akolmio}{{$[\blacktriangle]$}}
\newcommand{\bkolmio}{{$[\triangle]$}}
\begin{center}
\begin{tabular}{cllll}
\hline
                             & ${\rm IND}(\tau)=1$ & ${\rm IND}(\tau)=1$ & 
                               ${\rm IND}(\tau)=0$ & ${\rm IND}(\tau)=0$ \\
                             & ${\rm R}(\tau)=0$ & ${\rm R}(\tau)=1$ & 
                               ${\rm R}(\tau)=0$ & ${\rm R}(\tau)=1$ \\
\hline
                   $M(\tau)$ &  $n=39$ \anelio &   $n=14$ \bnelio &  $n=331$ \risti &  $n=101$ \risti \\
                   $A(\tau)$ &  $n=20$ \anelio &   $n=14$ \bnelio &  $n=172$ \risti &  $n=101$ \risti \\
                   $P(\tau)$ &  $n=20$ \anelio &   $n=14$ \bnelio &  $n=172$ \risti &  $n=101$ \risti \\
  $t_{\mathrm{min,1}}(\tau)$    &  $n=20$ \anelio &   $n=14$ \bnelio &  $n=172$ \risti &  $n=101$ \risti \\
  $t_{\mathrm{min,2}}(\tau)$    &  $n=5$ \akolmio &    $n=8$ \bkolmio&   $n=65$ \risti &   $n=47$ \risti \\
\hline
\end{tabular}
\end{center}
\noindent
The symbols in brackets are used in Fig. \ref{figurethree}, where
error bars are displayed only for ${\rm IND}(\tau)=1$ datasets.

CPS also gave the time scale of change of the light curve,
$T_{\mathrm{C}}(\tau)$, for all 370 reliable models.
The light curve of \object{V352~CMa} changed before the end of the segment only
in 37 datasets, where the $T_{\mathrm{C}}$ mean was 51 days. 
This mean exceeded 65 days in the other 333 datasets,
where the light curve did not change before the segment end.
Both values indicated that the applied dataset length,
$\Delta T_{\mathrm{max}}=30^{\mathrm{d}}$, gave reliable CPS results.

The light curves of independent datasets are shown in Fig. \ref{figuretwo}.
The continuous lines display the periodic curves ($K\ge1$).
The phases were first computed from 
$\phi_1=\mathrm{FRAC}[(t-t_{\mathrm{min,1}}(\tau))/P(\tau)]$,
where $\mathrm{FRAC}[x]$ removes the integer part of its argument $x$.
Then the phases $\phi_{\mathrm{al,1}}$ of the $t_{\mathrm{min,1}}(\tau)$ epochs
were computed from the constant period ephemeris 
HJD~$2451178.1245+7\fd158$E.
Finally, the data and the light curves were plotted as a function of
$\phi=\phi_1+\phi_{\mathrm{al,1}}$.
The dashed lines display the aperiodic curves ($K=0$), where
the ``phases'' are $\phi_i=(t_i-t_1)/(t_n-t_1)$. 
The SEG, SET, $\tau$ and $R(\tau)$ values are also given.

The long-term $M(\tau)$ and $A(\tau)$ changes of \object{V352~CMa} are
shown in Fig. \ref{figurethree}, panels a) and b).
PSM has been applied
to search for activity cycles in chromospheric Ca~II H\&K 
emission line data \citep[e.g.][]{Bal95}.
\citet{Rod00} applied this method to the following light curve parameters:
$M(\tau)$                   (axisymmetric part of spot distribution),
$A(\tau)$                   (non-axisymmetric part of spot distribution),
$M(\tau)-A(\tau)/2$         (minimum spotted area) or
$M(\tau)+A(\tau)/2$         (maximum spotted area).
We applied PSM to the same independent and reliable estimates of \object{V352~CMa}.
The false alarm probability of the best cycle, 
$P_{\mathrm{C}}=11\fy7 \pm 0\fy5$ for the $n=39$ values
of $M(\tau)$, was $F=0.01$.
The cycles for the $n=20$ values of $A(\tau)$, 
$M(\tau)-A(\tau)/2$ and $M(\tau)+A(\tau)/2$
reached only $F \ge 0.26$.

The weighted mean of the independent and 
reliable $P(\tau)$ $(n=20)$ estimates
was $P_{\mathrm{w}} \pm \Delta P_{\mathrm{w}} = 7\fd24 \pm 0\fd22$.
This range of period changes, $P_{\mathrm{w}} \pm 3 \Delta P_{\mathrm{w}}$,
(Fig. \ref{figurethree}c: dotted lines) 
gave $Z=6\Delta P_{\mathrm{w}}/P_{\mathrm{w}}=0.19\equiv 19$ \% 
\citep[][Eq. 14]{Leh11}. 
The average of the light curve half amplitude $
A(\tau)/2$ in these datasets was only
$0.011$. 
The accuracy of the photometry, $\sigma_{\mathrm{N}}=0.007$,
gave an amplitude to noise ratio of $A/N \la 2$. 
Even if the spurious period changes were $Z_{\mathrm{spu}} \ge 0.15$,
the estimated real physical changes could be
$Z_{\mathrm{phys}}=(Z^2 - Z_{\mathrm{spu}}^2)^{-1/2} \approx 0.12$
\citep[][Table 3 and Eq. 15]{Leh11}.
We assumed that the solar law of surface differential rotation could be
applied to \object{V352~CMa} and that the period $P(\tau)$ changes could be used
to trace surface differential rotation.
We used $k\approx Z_{\mathrm{phys}}/h$, 
where the minimum and maximum latitudes of spot activity 
were $b_{\mathrm{min}}$ and  $b_{\mathrm{max}}$, and
$h=\sin^2{b_{\mathrm{max}}}-\sin^2{b_{\mathrm{min}}}$
\citep{Jet00}. 
The maximum value, $h=1$, would be reached, 
if spots were formed at all latitudes between the equator 
and pole of \object{V352~CMa}.
All other alternatives give 
$h < 1 \Rightarrow |k| > Z_{\mathrm{phys}} \approx 0.12$. 
This result was comparable to the solar value $k=0.20$, 
because sunspots form at latitudes
$\pm 30^{\mathrm{o}}$ (i.e. $h=0.25$).

We applied the non-weighted Kuiper test \citep{Jet96} to the
reliable primary minima $t_{\mathrm{min,1}}(\tau) ~(n=20)$ of independent datasets. 
The test range was between
$0.85 P_\mathrm{w} = 6\fd0$ and $1.15 P_\mathrm{w} = 8\fd1$.
The critical level of the best period 
$P_{\mathrm{al,1}}=7\fd158 \pm 0\fd002$ was $Q=0.004$.
The test for the reliable 
$t_{\mathrm{min,1}}(\tau)$ and $t_{\mathrm{min,2}}(\tau)$  $(n=25)$ 
of all independent datasets gave the same result,
$P_{\mathrm{al,1,2}}=7\fd158 \pm 0\fd002$,
but the critical level of this periodicity was lower, $Q=0.056$.
The phases of all primary and secondary minima are shown in Fig. \ref{figurethree}d
using the ephemeris HJD~$2451178\fd1245+7\fd158$E.
The phases of $t_{\mathrm{min,1}}$ estimates were very stable 
between 1998 and 2009 (Fig. \ref{figurethree}d: Filled squares). 
However, this long--lived structure vanished after 2009 when the light curve 
amplitudes $A(\tau)$ decreased. 

We then applied PSM to all datasets analysed with CPS.
The abbreviation ``C$_0$'' was used for the
cases: $ P_{\mathrm{CPS}} - \sigma_{P_{\mathrm{CPS}}} \le 
            P_{\mathrm{PSM}} 
         \le P_{\mathrm{CPS}} + \sigma_{P_{\mathrm{CPS}}}$.
Our results were
\begin{center}
\begin{tabular}{clll}
Criterion          & $K=1$ or 2 $~[\mathrm{\%}]$& $K = 1 ~[\mathrm{\%}]$ & $K=2 ~[\mathrm{\%}]$ \\
\hline
 $A(\tau)>      0 $ & $173/307\equiv 56    $ & $141/170\equiv 83    $ & $ 32/137\equiv 23    $ \\ 
 $A(\tau)>  0.015 $ & $129/240\equiv 54    $ & $ 97/112\equiv 87    $ & $ 32/128\equiv 25    $ \\ 
 $A(\tau)>  0.030 $ & $ 15/ 40\equiv 38    $ & $  3/  3\equiv100    $ & $ 12/ 37\equiv 32    $ \\ 
\hline
   $F <     0.5  $ & $152/258\equiv 59    $ & $126/149\equiv 85    $ & $ 26/109\equiv 24    $ \\ 
   $F <     0.1  $ & $ 52/ 86\equiv 60    $ & $ 41/ 42\equiv 98    $ & $ 11/ 44\equiv 25    $ \\ 
   $F <     0.05 $ & $ 25/ 36\equiv 69    $ & $ 18/ 18\equiv100    $ & $  7/ 18\equiv 39    $ \\ 
\hline
\end{tabular}
\end{center}
Case $C_0$ was true only in 56\% cases for all periodicity detections 
($n=307: K=1$ or 2, $A(\tau)>0.00$). This occurred more often for $K=1$ (83\%) 
than for $K=2$ (23\%) models. 
These fractions increased for higher amplitude, $A(\tau)>0.015$, light curves.
For the highest amplitudes, $A(\tau) > 0.030$,  
case $C_0$ was true for all $K=1$ models (100\%), but not for all $K=2$ models (32\%).
As expected, the probability for case $C_0$ being true also increased when 
the false alarm probability $F$ decreased.
For $F<0.05$, case $C_0$ was true for all $K=1$ models (100\%), 
but again not for all $K=2$ models (39\%).
In conclusion, PSM did not always detect the correct period for the $K=2$
light curves. This occurred also for high amplitudes $A(\tau)$
or for small false alarm probabilities $F$.

\section{Conclusions}

We wanted to test the performance of CPS and applied it to 
the low amplitude light curves of \object{V352~CMa}. 
These data were challenging, because the dispersions of 
the $\Delta V_{\mathrm{V-C}}$ and  $\Delta V_{\mathrm{K-C}}$ magnitudes
were comparable (Figs. \ref{figureone}ab). 
In a total of 485 datasets, CPS detected
no periodicity in 178 $\equiv$ 37\% datasets. 
Comparison of CPS and PSM revealed that both methods gave the same best
periods for high amplitude to noise ratio light curves, 
{\it but only if} the correct model for the data was a sinusoid $(K=1)$.
The best periods were {\it mostly} not 
the same for the $K=2$ models --
not even for the higher $A/N$ light curves.
We conclude that 
if the period $P(\tau)$ or the model order $K$ are not correct, 
then the $M(\tau)$, $A(\tau)$,
$t_{\mathrm{min,1}}$ and $t_{\mathrm{min,2}}$ estimates are 
in many cases also not correct. 

We detected signs of an activity cycle, 
$P_{\mathrm{C}}=11\fy7 \pm 0\fy5 ~(F=0.010)$, 
in the mean $M(\tau) ~(n=39)$. 
However, it was only $1^{\mathrm{y}}$ shorter 
than the time span of the data, $\Delta T=13\fy0$.
The $P(\tau)$ period changes could be mostly 
spurious $(Z_{\mathrm{spu}} > 0.15)$ 
due to low $A/N$ ratio \citep[][Table 3]{Leh11}.
If the law of solar differential rotation were valid in \object{V352~CMa}, 
and these $P(\tau)$ changes traced this phenomenon,
the surface differential rotation coefficient would be 
$|k| > Z_{\mathrm{phys}} \approx 0.12$.
The Kuiper method detected an active longitude rotating with a period of 
$P_{\mathrm{al,1}}=7\fd158 \pm 0\fd002 ~(Q=0.004)$.
This long--lived structure was present between 1998 and 2009, but vanished
when the amplitudes of the light curves fell close to zero in 2010.

\begin{acknowledgements}
This research at the Department of Physics (University of Helsinki)
was performed in collaboration with the participants
of the course ``Variable stars'', which was lectured in spring 2012. 
This work has made use of the SIMBAD database at CDS, Strasbourg,
France and NASA's Astrophysics Data System (ADS) bibliographic services.
The work by PK and JL was supported by the Vilho, Yrj\"{o} and Kalle V\"{a}is\"{a}l\"{a} 
Foundation. The automated astronomy program at Tennessee State University 
has been supported by NASA, NSF, TSU and the State of Tennessee 
through the Centers of Excellence program.
\end{acknowledgements}

\bibliographystyle{aa}
\bibliography{v352cmareferences}
\end{document}